\documentclass{article}

\usepackage{arxiv}

\usepackage[utf8]{inputenc} 
\usepackage[T1]{fontenc}    
\usepackage{hyperref}       
\usepackage{url}            
\usepackage{booktabs}       
\usepackage{amsfonts}       
\usepackage{nicefrac}       
\usepackage{microtype}      
\usepackage{lipsum}		
\usepackage{graphicx}
\usepackage[numbers]{natbib}
\usepackage{doi}
\usepackage{float}

\usepackage{pifont}
\usepackage{amsmath}

\usepackage{listings}

\usepackage{xcolor}
\definecolor{codegreen}{rgb}{0,0.6,0}
\definecolor{codegray}{rgb}{0.5,0.5,0.5}
\definecolor{codepurple}{rgb}{0.58,0,0.82}
\definecolor{backcolour}{rgb}{1,1,1}
\lstdefinestyle{mystyle}{
    backgroundcolor=\color{backcolour},   
    commentstyle=\color{codegreen},
    keywordstyle=\color{magenta},
    numberstyle=\tiny\color{codegray},
    stringstyle=\color{codepurple},
    basicstyle=\ttfamily\footnotesize,
    breakatwhitespace=false,         
    breaklines=true,                 
    captionpos=t, 
    keepspaces=true,                 
    numbers=none, 
    numbersep=5pt,                  
    showspaces=false,                
    showstringspaces=false,
    showtabs=false,                  
    tabsize=2
}
\makeatletter
\pretocmd\lst@makecaption{\noindent{\rule{\linewidth}{1pt}}}{}{}
\makeatother

\usepackage{rotating}

\title{LightDSA: A Python-Based Hybrid Digital Signature Library and Performance Analysis of RSA, DSA, ECDSA and EdDSA in Variable Configurations, Elliptic Curve Forms and Curves}

\author{ \href{https://orcid.org/0000-0002-0345-0088}{\includegraphics[scale=0.06]{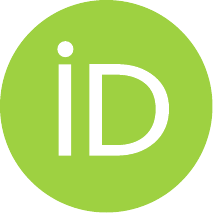}\hspace{1mm}Sefik Serengil} \\
	Solution Engineering \\
	Vorboss Limited\\
        \texttt{London, UK} \\
	\texttt{sefik.serengil@vorboss.com} \\
	\And
	\href{https://orcid.org/0000-0003-1250-5949}{\includegraphics[scale=0.06]{orcid.pdf}\hspace{1mm}Alper Ozpinar} \\
	Department of Management\\
	Ibn Haldun University\\
        \texttt{Istanbul, TURKIYE} \\
	\texttt{alper@ozpinar.org} \\
}

\renewcommand{\shorttitle}{LightDSA}

\hypersetup{
    pdftitle={LightDSA: A Python-Based Hybrid Digital Signature Library and Performance Analysis of RSA, DSA, ECDSA and EdDSA in Variable Configurations},
    pdfsubject={cs.CR},
    pdfauthor={Sefik Serengil, Alper Ozpinar},
    pdfkeywords={Cryptography, Digital Signature, ECDSA, EdDSA, RSA, Python},
}

\begin{document}
\maketitle

\begin{abstract}
Digital signature algorithms (DSAs) are fundamental to cryptographic security, ensuring data integrity and authentication. While RSA, DSA, ECDSA, and EdDSA are widely used, their performance varies significantly depending on key sizes, hash functions, and elliptic curve configurations. In this paper, we introduce LightDSA, a hybrid and configurable digital signature library that supports RSA, DSA, ECDSA, and EdDSA with flexible form and curve selection, open sourced at \url{https://github.com/serengil/LightDSA}. Unlike conventional implementations that impose strict curve-form mappings—such as Weierstrass for ECDSA and Edwards for EdDSA—LightDSA allows arbitrary combinations, enabling a broader performance evaluation. We analyze the computational efficiency of these algorithms across various configurations, comparing key generation, signing, and verification times. Our results provide insights into the trade-offs between security and efficiency, guiding the selection of optimal configurations for different cryptographic needs.
\end{abstract}


\section{Introduction}

Digital signatures are a cornerstone of modern cryptography, providing a mechanism to verify the authenticity and integrity of digital messages and documents. Widely used in various security protocols, digital signature algorithms (DSAs) ensure that the data has not been tampered with and that it originates from a trusted sender. Among the most commonly used digital signature algorithms are RSA \cite{rsa}, DSA \cite{dsa}, ECDSA \cite{ecdsa}, and EdDSA \cite{eddsa}, each offering unique strengths and trade-offs in terms of security, efficiency, and flexibility. These algorithms are essential in securing communications, enabling digital certificates, and facilitating blockchain transactions, among other applications.

Despite their widespread adoption, the performance of these algorithms varies significantly depending on several factors, including the choice of key size, elliptic curve type, and hash function. RSA has been the standard for digital signatures for decades, relying on large integer factorization for its security. However, its performance decreases as the key size increases, making it less efficient for modern applications that require smaller, faster cryptographic operations. 

Elliptic Curve Cryptography (ECC) \cite{miller1985use} \cite{koblitz} is a branch of public-key cryptography based on the algebraic structure of elliptic curves over finite fields. ECC offers high levels of security with relatively small key sizes, making it highly efficient compared to traditional systems like RSA. This efficiency has led to its widespread adoption in modern cryptographic protocols. While ECC is widely used for key exchange protocols and digital signatures, it can also be applied to encryption schemes, such as EC ElGamal \cite{ecelgamal}. However, despite its theoretical potential for encryption and decryption, EC ElGamal's decryption process requires solving the Elliptic Curve Discrete Logarithm Problem (ECDLP), which is computationally difficult and impractical for real-world applications. This makes EC ElGamal unsuitable for most encryption purposes \cite{serengil2024lightphe} \cite{serengil2024lightphe2}. In contrast, Elliptic Curve Diffie-Hellman (ECDH) \cite{bernstein2006curve25519}, a key exchange protocol based on ECC, is highly efficient and practical for establishing secure communication channels. Due to its speed and strong security guarantees, ECDH has gained significant adoption, with tools like GPG incorporating ECDH support for encryption in recent versions. Similarly, Elliptic Curve Digital Signature Algorithm (ECDSA) and Edwards-curve Digital Signature Algorithm (EdDSA) are two widely adopted and efficient elliptic curve-based digital signature algorithms. These algorithms provide strong security with relatively small key sizes, making them ideal choices for digital signature applications in environments requiring fast and scalable cryptographic operations.

ECDSA (Elliptic Curve Digital Signature Algorithm) and EdDSA (Edwards-curve Digital Signature Algorithm) also address RSA's issue by leveraging elliptic curve cryptography (ECC), which provides strong security with smaller key sizes. While ECDSA is based on the Weierstrass curve form, EdDSA employs the Edwards curve form \cite{bernstein2015eddsa}, known for its efficiency and resistance to certain types of attacks \cite{bernstein2007faster}. Both elliptic curve-based algorithms have gained significant traction in cryptographic standards and protocols.

However, despite the popularity of these algorithms, many implementations follow fixed configurations, such as specific elliptic curve forms or pre-defined curves. ECDSA, for example, is typically implemented using the secp256k1 or secp256r1 (p256) curve, while EdDSA often uses Ed25519 \cite{ed25519}. While these defaults are widely accepted, they may not always represent the best choice for every use case \cite{bernstein2013safecurves}.

The main goal of this paper is to present LightDSA, a Python-based digital signature library that supports RSA, DSA, ECDSA, and EdDSA, open sourced at \url{https://github.com/serengil/LightDSA}. The library offers flexibility in terms of elliptic curve forms (Weierstrass \cite{weierstrass}, Koblitz \cite{ecbinary}, and Edwards \cite{edwards} \cite{bernstein2008twisted}) and supports a wide range of pre-defined curves such as secp256k1, secp256r1 or ed25519, allowing users to experiment with different configurations. The library automatically determines the appropriate hash function based on the key size or curve order, supporting common cryptographic hash functions like SHA-1, SHA-224, SHA-256, SHA-384, and SHA-512 \cite{sha512}.

In this paper, we provide an extensive performance evaluation of LightDSA, comparing RSA, ECDSA, and EdDSA across different configurations. Specifically, we analyze the impact of curve selection, hash function choice, and key size on the performance of digital signature operations, including key generation, signing, and verification. We also explore the trade-offs between security and efficiency, highlighting the suitability of various configurations for different cryptographic needs.

The findings presented in this paper aim to guide cryptographic practitioners and researchers in selecting the most appropriate digital signature algorithm and configuration for their specific use cases. By providing a highly configurable and performance-optimized framework, LightDSA offers a tool to better understand the strengths and weaknesses of various digital signature approaches, contributing to the ongoing development of more efficient cryptographic systems.

\section{Digital Signature Algorithms}

In this section, we will focus on three widely used cryptographic methods for ensuring the authenticity and integrity of digital messages: RSA, ECDSA, and EdDSA. RSA, one of the oldest and most widely adopted algorithms, uses a pair of keys—private and public—for signing and verification. It provides strong security based on the difficulty of factoring large numbers. ECDSA, which uses elliptic curve cryptography, offers similar security to RSA but with much smaller key sizes, making it more efficient and suitable for resource-constrained environments. Finally, EdDSA, an enhanced version of ECDSA, uses a different elliptic curve and focuses on improving performance, security, and resistance to side-channel attacks. Each of these algorithms plays a crucial role in digital signatures, offering varying benefits depending on the specific requirements of the application.


\subsection{RSA}

RSA is a widely-used public-key cryptosystem that operates on two key principles: encryption and decryption. In the context of homomorphic encryption and secure message transfer, the public key is used for encryption, and the private key is used for decryption \cite{phe}. On the other hand, in the context of digital signatures, the private key is used for signing, while the public key is used for verification \cite{rsa_for_signature}.

To sign a message, the sender uses their private key, which is represented as d, to generate the signature. The hashed message m is raised to the power of d and then taken modulo n, as described in Equation \ref{eq:rsa_encrypt}.

\begin{equation}
\label{eq:rsa_encrypt}
    {c} = { (m)^d \quad mod \quad n }
\end{equation}

Here, c is the resulting signature. This process ensures that only the holder of the private key can generate a valid signature. To verify the signature, the recipient uses the sender's public key e. The signature c is raised to the power of e and taken modulo n, as shown in Equation \ref{eq:rsa_decrypt}. If the result matches the hash of the original message, the signature is valid, proving both the authenticity of the message and that it has not been tampered with. 

\begin{equation}
\label{eq:rsa_decrypt}
    {m} = { (c)^e \quad mod \quad n }
\end{equation}

The relationship between the public and private keys is governed by the Equation \ref{eq:rsa_theory_1} and \ref{eq:rsa_theory_1_1}.

\begin{equation}
\label{eq:rsa_theory_1}
    {e \times d} = { 1 \mod \phi(n)}
\end{equation}

\begin{equation}
\label{eq:rsa_theory_1_1}
    {e \times d} = { k \times \phi(n) + 1}
\end{equation}

If this rule is applied the encryption - decryption operations respectively

\begin{equation}
\label{eq:rsa_proof_1}
    {(m ^ d) ^ e} \mod n
\end{equation}

\begin{equation}
\label{eq:rsa_proof_2_0}
    {m ^ {k \times \phi(n) + 1}} \mod n
\end{equation}

\begin{equation}
\label{eq:rsa_proof_3}
    {m^{k \times \phi(n)} \times m} \mod n
\end{equation}

\begin{equation}
\label{eq:rsa_proof_4}
    {(m^{\phi(n)})^k \times m} \mod n
\end{equation}

According to Fermat-Euler Theorem, raising any integer a, which is coprime to n, to the power of Euler's totient function results in 1 modulo n. Put Fermat-Euler theorem back to the equation above as follows

\begin{equation}
\label{eq:rsa_proof_2}
    {(m^{\phi(n)})^k \times m} \mod n
\end{equation}

\begin{equation}
\label{eq:rsa_proof_2_2}
    {(1)^k \times m} \mod n
\end{equation}

\begin{equation}
\label{eq:rsa_proof_2_3}
    m \mod n
\end{equation}

In summary, applying the private key first and then the public key will recover the original plaintext message as shown in Equation \ref{eq:rsa_proof_3_3}.

\begin{equation}
\label{eq:rsa_proof_3_3}
    {(m ^ d) ^ e} \mod n = m \mod n
\end{equation}

This ensures that the encryption and decryption operations are mathematically linked, allowing for secure digital signatures. Thus, RSA provides a robust mechanism for ensuring message integrity and authenticity in digital communication, where the private key signs the message, and the public key verifies it.


\subsection{DSA}

DSA is a modified version of the Schnorr and ElGamal \cite{elgamal} digital signature algorithms. It requires the selection of two prime number p and q that satisfy the following conditions:

\begin{equation}
\label{eq:dsa_1}
    p - 1 \mod q = 0
\end{equation}

A generator g is then generated randomly using, where h is chosen from the range [2, p-2]

\begin{equation}
\label{eq:dsa_2}
    g \mod p= h ^ {\frac{p-1}{q}} \mod p
\end{equation}

If we raise both sides to power of q as

\begin{equation}
\label{eq:dsa_3}
    g^q \mod p = {(h ^ {\frac{p-1}{q}})}^q \mod p
\end{equation}

\begin{equation}
\label{eq:dsa_4}
    g^q \mod p = h ^ {p-1} \mod p
\end{equation}

According to the Fermat's Little Theorem, right hand side must be equal to 1.

\begin{equation}
\label{eq:dsa_5}
    g^q \mod p = 1
\end{equation}

The signer then selects a private key x in the range [1, q-1] and computes the corresponding public key y as follows:

\begin{equation}
\label{eq:dsa_6}
    y = g ^ x \mod p
\end{equation}

The signer then selects a random integer k for each signature, chosen from the range [1, q-1], and the signature (r, s) pair is calculated as follows:

\begin{equation}
\label{eq:dsa_7}
    r = (g^k \mod p) \mod q
\end{equation}

\begin{equation}
\label{eq:dsa_8}
    s = \frac{H(m) + x \times r}{k} \mod q
\end{equation}

In verification process, the verifier calculates $u_1$ and $u_2$ as follows:

\begin{equation}
\label{eq:dsa_9}
    u_1 = \frac{H(m)}{s} \mod q
\end{equation}

\begin{equation}
\label{eq:dsa_10}
    u_2 = \frac{r}{s} \mod q
\end{equation}

The verifier then checks whether the following calculations are equal to the r part of the signature.

\begin{equation}
\label{eq:dsa_11}
    ((g^{u_1} \mod p) \times (y^{u_2} \mod p)) \mod q = r
\end{equation}

To prove the correctness of the verification process, recall the computation of s

\begin{equation}
\label{eq:dsa_12}
    s = \frac{H(m) + x \times r}{k} \mod q
\end{equation}

If we find k from that equation as follows

\begin{equation}
\label{eq:dsa_13}
    k = \frac{H(m) + x \times r}{s} \mod q
\end{equation}

Move divisor s to the additions in the dividend

\begin{equation}
\label{eq:dsa_14}
    k = \frac{H(m)}{s} + \frac{x \times r}{s} \mod q
\end{equation}

Move this equation to the exponents with the same base g and the same modulo p.

\begin{equation}
\label{eq:dsa_15}
    g^k \mod p= g^{(\frac{H(m)}{s} + \frac{x \times r}{s} \mod q)} \mod p 
\end{equation}

According to the product rule of exponents, this can be re-arranged as

\begin{equation}
\label{eq:dsa_16}
    g^k \mod p = g^{(\frac{H(m)}{s} \mod q)} \times g^{(\frac{x \times r}{s} \mod q)} \mod p
\end{equation}

According to the power of power rule, this can be represented as follows:

\begin{equation}
\label{eq:dsa_17}
    g^k \mod p = g^{(\frac{H(m)}{s} \mod q)} \times ({g^x})^{(\frac{r}{s} \mod q)} \mod p
\end{equation}

The base g raised to the power of the private key x is equal to the public key y.

\begin{equation}
\label{eq:dsa_18}
    g^k \mod p = g^{(\frac{H(m)}{s} \mod q)} \times y^{(\frac{r}{s} \mod q)} \mod p
\end{equation}

We can replace the exponents with $u_1$ and $u_2$.

\begin{equation}
\label{eq:dsa_19}
    g^k \mod p = g^{u_1} \times y^{u_2} \mod p
\end{equation}

Thus, the verifier calculates $g^k$ during verification, where k is the random key generated by the signer, and the verifier does not know this value directly.

Meanwhile, the signer calculates r as $g^k \mod p$, and then takes the result modulo q.

\begin{equation}
\label{eq:dsa_20}
    (g^k \mod p) \mod q = (g^{u_1} \times y^{u_2} \mod p) \mod q
\end{equation}

\begin{equation}
\label{eq:dsa_21}
    r = (g^{u_1} \times y^{u_2} \mod p) \mod q
\end{equation}


\subsection{ECDSA}

A user should select a random number $k_a$ as their private key. Then, they compute the corresponding point on the elliptic curve by multiplying the private key with the base point G, obtaining their public key $Q_a$.

\begin{equation}
\label{eq:ecdsa_1}
    Q_a = k_a \times G
\end{equation}

Then, a random point R on the elliptic curve should be selected.

\begin{equation}
\label{eq:ecdsa_2}
    R = k_r \times G
\end{equation}

The private key should be chosen only once, but a new random point must be generated for each signature. If the same random point is reused across multiple signatures, an attacker can exploit this to recover the private key by solving for it algebraically. This is a critical security requirement in elliptic curve cryptography to prevent key leakage.

The signature (r, s) pair should be calculated as

\begin{equation}
\label{eq:ecdsa_3_r}
    r = R_x
\end{equation}

\begin{equation}
\label{eq:ecdsa_3_s}
    s = \frac{m + r \times k_a}{k_r} \mod n
\end{equation}

In verification, user should calculate the pair $u_1$ and $u_2$ as

\begin{equation}
\label{eq:ecdsa_4}
    u_1 = \frac{m}{s} \times G
\end{equation}

\begin{equation}
\label{eq:ecdsa_5}
    u_2 = \frac{r}{s} \times Q_a
\end{equation}

Finally, user should find the addition of the points $u_1$ and $u_2$ and check its x coordinate is equal to r. If this condition is true, then signature is valid.

\begin{equation}
\label{eq:ecdsa_6}
    (u_1 + u_2)_x = r
\end{equation}

To prove this scheme, we need to replace $u_1$ and $u_2$ first.

\begin{equation}
\label{eq:ecdsa_6_2}
    u_1 + u_2 = \frac{m}{s} \times G + \frac{r}{s} \times Q_a
\end{equation}

Public key $Q_a$ can be expressed as $k_a \times G$

\begin{equation}
\label{eq:ecdsa_7}
    u_1 + u_2 = \frac{m}{s} \times G + \frac{r}{s} \times k_a \times G
\end{equation}

Both terms in the addition have common s divisor and base point G multiplier. 

\begin{equation}
\label{eq:ecdsa_8}
    u_1 + u_2 = \frac{m + r \times k_a}{s} \times G
\end{equation}

Replace the s in the divisor from Equation \ref{eq:ecdsa_3_s}

\begin{equation}
\label{eq:ecdsa_9}
    u_1 + u_2 = \frac{m + r \times k_a}{\frac{m + r \times k_a}{k_r}} \times G
\end{equation}

Dividends can be simplified and the addition of $u_1$ and $u_2$ becomes random point R.

\begin{equation}
\label{eq:ecdsa_10}
    u_1 + u_2 = k_r \times G = R
\end{equation}

On the other hand, we checked the equality of x coordinate $u_1$ and $u_2$ addition to r part of the signature where r was the x coordinate of random point already.

\subsection{EdDSA}

Similar to ECDSA, user should generate a random integer once as their private key and compute its corresponding point by multiplying their private key with base point G.

\begin{equation}
\label{eq:eddsa_public}
    Q_a = k_a \times G
\end{equation}

Secondly, the user should generate a random point by hashing the message. This ensures its randomness and prevents the leakage of the same random key for different signatures.

\begin{equation}
\label{eq:eddsa_1}
    r = H(H(m) + m)
\end{equation}

Then, calculates the corresponding random point by multiplying random number with the base point G.

\begin{equation}
\label{eq:eddsa_2}
    R = r \times G
\end{equation}

Threafter, the x coordinate of random point, the x coordinate of public key and message should be summed to calculate h, where p is the prime over which the elliptic curve is defined.

\begin{equation}
\label{eq:eddsa_3}
    h = R_x + (Q_a)_x + message \mod p
\end{equation}

Finally, random number derived from message and h times private key $k_a$ should be summed to calcualte s.

\begin{equation}
\label{eq:eddsa_4}
    s = r + h \times k_a
\end{equation}

The (R, s) pair will be dispatched as a signature, where R is a point on the elliptic curve with x and y coordinates, and s is an integer.

Once message and signature (R, s) dispatched, verifier can compute same h as

\begin{equation}
\label{eq:eddsa_3_2}
    h = R_x + (Q_a)_x + message \mod p
\end{equation}

Then, verifier will compute $P_1$ and $P_2$ as

\begin{equation}
\label{eq:eddsa_4_2}
    P_1 = s \times G
\end{equation}

\begin{equation}
\label{eq:eddsa_5}
    P_2 = R + h \times Q_a
\end{equation}

Signature is valid only if $P_1$ and $P_2$ are equal.

\begin{equation}
\label{eq:eddsa_verify}
    P_1 = P_2
\end{equation}

To prove the correctness of EdDSA schema, put s back to $P_1$ computation

\begin{equation}
\label{eq:eddsa_4_3}
    P_1 = s \times G = (r + h \times k_a) \times G = r \times G + h \times k_a \times G
\end{equation}

In the equation above, $r \times G$ refers to random point R, and $k_a \times G$ denotes public key $Q_a$. That is exactly equal to $P_2$. So, this shows the correctness of the verification process of EdDSA.

\begin{equation}
\label{eq:eddsa_5_2}
    P_1  = R + h \times Q_a = P_2
\end{equation}


\section{Python Library}

In the previous section, we focused on the theory behind three major digital signature algorithms. LightDSA aims to provide a simple and intuitive interface for building cryptosystems using these algorithms. Users do not need to delve into the underlying processes. To construct a cryptosystem, they can simply initialize a LightDSA object and set the algorithm to RSA, DSA, ECDSA, or EdDSA. For elliptic curve-based algorithms, users can optionally choose the curve type, such as Weierstrass, Koblitz, or Edwards, and specify the curve, as demonstrated in the following snippet. Once set up, any message can be signed using the sign method of the LightDSA object.

\begin{minipage}{\linewidth} 
\begin{lstlisting}[frame=tb, caption=Building a Cryptosystem and Signing a Message, label=building_cs, language=Python]
# !pip install lightdsa
from lightdsa import LightDSA

# construct a cryptosystem
dsa = LightDSA(
   algorithm_name = "eddsa", # or ecdsa, rsa, dsa
   form_name = "edwards", # or weierstrass, koblitz
   curve_name = "ed25519",
)

# export public key
dsa.export_keys("public.txt", public = True)

# sign a message
message = "Hello, world!"
signature = dsa.sign(message)
\end{lstlisting}
\end{minipage}
\newline
\newline

On the verifier side, the LightDSA object should be initialized with the same configuration, along with the exported public key. The verify method can then be used to verify a message and its corresponding signature. If the signature is invalid, the method will raise an error.

\begin{minipage}{\linewidth} 
\begin{lstlisting}[frame=tb, caption=Veriying a Message, label=verify, language=Python]
# construct the cryptosystem with public key
verifier_dsa = LightDSA(
   algorithm_name = "eddsa", # or ecdsa, rsa, dsa
   form_name = "edwards", # or weierstrass, koblitz
   curve_name = "ed25519",
   key_file = "public.txt",
)

assert verifier_dsa.verify(message, signature) is True
\end{lstlisting}
\end{minipage}
\newline
\newline

\section{Experiments}

Table \ref{tab:rsa} presents the performance of the RSA algorithm with different key sizes and hash functions, detailing the key generation time, signing time, and verification time for each configuration. With a 1024-bit key and SHA-160 hash, RSA key generation takes 0.0480 seconds, while signing and verification operations complete in 0.0031 and 0.0034 seconds, respectively. As the key size increases to 2048 bits with SHA-224, key generation time rises to 0.5314 seconds, and signing and verification times increase to 0.0221 and 0.0219 seconds, respectively. At 3072 bits with SHA-256, key generation time grows significantly to 9.3434 seconds, with signing and verification times of 0.0704 and 0.0693 seconds. The trend continues with a 7680-bit key and SHA-384, where key generation takes 43.8724 seconds, while signing and verification times increase to 0.9308 and 0.9995 seconds, respectively. Finally, at 15360 bits with SHA-512, RSA key generation time surges to 455.00 seconds, while signing and verification times rise drastically to 7.0423 and 7.6164 seconds. These results demonstrate that as the RSA key size increases, the computational overhead grows exponentially, making RSA increasingly impractical for larger key sizes, particularly in applications requiring frequent key generation.

\begin{table}[h!]
\centering
\caption{Performance of RSA with different key sizes and hashes}
\begin{tabular}{lll|lll}
\hline
\textbf{Algorithm} & \textbf{Key Size} & \textbf{Hash} & \textbf{KeyGen (s)} & \textbf{Sign (s)} & \textbf{Verify (s)} \\
\hline
RSA & 15360 & sha512 & 455.00 & 7.0423 & 7.6164 \\
RSA & 7680  & sha384 & 43.8724 & 0.9308 & 0.9995 \\
RSA & 3072  & sha256 & 9.3434 & 0.0704 & 0.0693 \\
RSA & 2048  & sha224 & 0.5314 & 0.0221 & 0.0219 \\
RSA & 1024  & sha160 & 0.0480 & 0.0031 & 0.0034 \\
\hline
\end{tabular}
\label{tab:rsa}
\end{table}

The table \ref{tab:dsa} presents the performance of the Digital Signature Algorithm (DSA) with varying key sizes and hash functions. It shows the time taken for key generation, signing, and verification for different configurations. The key sizes range from 1024 bits to 15360 bits, with corresponding hash functions including sha1, sha224, sha256, sha384, and sha512. As the key size increases, the time required for key generation grows significantly, with the 15360-bit key taking the longest at 7172.1030 seconds. However, the signing and verification times remain relatively low across all configurations, with the fastest being 0.0006 seconds for signing and 0.0012 seconds for verification at a 1024-bit key with the sha160 hash. The results highlight the trade-off between key size and computational efficiency, with larger keys offering higher security but requiring more time for key generation.

\begin{table}[h!]
\centering
\caption{Performance of DSA with different key sizes and hashes}
\begin{tabular}{lll|lll}
\hline
\textbf{Algorithm} & \textbf{Key Size} & \textbf{Hash} & \textbf{KeyGen (s)} & \textbf{Sign (s)} & \textbf{Verify (s)} \\
\hline
DSA       & 15360      & sha512 & 7172.1030& 0.1937 & 0.3868 \\
DSA       & 7680       & sha384 & 432.1513 & 0.0412 & 0.0833 \\
DSA       & 3072       & sha256 & 41.2033  & 0.0065 & 0.0126 \\
DSA       & 2048       & sha224 & 3.9034   & 0.0033 & 0.0062 \\
DSA       & 1024       & sha160   & 0.8196   & 0.0006 & 0.0012 \\
\hline
\end{tabular}
\label{tab:dsa}
\end{table}

Table \ref{tab:edwards} compares the performance of Edwards curve cryptography for ECDSA and EdDSA, showing key generation, signing, and verification times across different curve sizes. Both schemes offer efficient performance, with execution times increasing gradually as the security level grows. Larger curves like e521 and ed448 maintain practical performance, while smaller curves like ed25519 provide the fastest execution. EdDSA generally outperforms ECDSA in signing operations, and both schemes scale more efficiently than RSA. Overall, Edwards curves are well-suited for fast and secure digital signatures.

\begin{table}[H]
\centering
\caption{Performance of Prime Field Edwards Form for ECDSA and EdDSA}
\begin{tabular}{lll|lll|lll}
\hline
 & & &  & \textbf{EdDSA} & & & \textbf{ECDSA} & \\
 \hline
\textbf{Curve} & \textbf{n} & \textbf{Hash} & \textbf{KeyGen (s)} & \textbf{Sign (s)} & \textbf{Verify (s)} & \textbf{KeyGen (s)} & \textbf{Sign (s)} & \textbf{Verify (s)} \\
\hline
e521 \cite{e521}                                 & 519 & sha512 & 0.0667  & 0.032  & 0.1357 & 0.0656  & 0.0647 & 0.1368 \\
rfc5832-512 & 510 & sha512 & 0.062   & 0.0314 & 0.1204 & 0.063   & 0.0631 & 0.1242 \\
numsp512t1                           & 510 & sha512 & 0.0585  & 0.0321 & 0.1253 & 0.064   & 0.0626 & 0.1433 \\
ed448 \cite{ed448}                                & 446 & sha512 & 0.0436  & 0.0259 & 0.0919 & 0.0466  & 0.0497 & 0.0992 \\
ed448godilocks \cite{ed448godilocks} & 446 & sha512 & 0.0516 & 0.0376 & 0.1131 & 0.0457 & 0.0551 & 0.0899 \\ 
curve41417 \cite{curve41417}                           & 411 & sha512 & 0.0381  & 0.0235 & 0.0728 & 0.0528  & 0.0363 & 0.0976 \\
numsp384t1                           & 382 & sha384 & 0.0356  & 0.0201 & 0.0616 & 0.0296  & 0.0301 & 0.0688 \\
e382 \cite{e521} & 380 & sha384 & 0.0347 & 0.0272 & 0.1340 & 0.0304 & 0.0328 & 0.0699 \\ 
rfc5832-256 & 255 & sha256 & 0.0125  & 0.0122 & 0.0256 & 0.0129  & 0.0132 & 0.0279 \\
mdc201601 \cite{milliondollarcurve} & 254 & sha256 & 0.0118  & 0.012  & 0.0244 & 0.0112  & 0.0114 & 0.0259 \\
numsp256t1                           & 254 & sha256 & 0.012   & 0.0122 & 0.0369 & 0.0119  & 0.0117 & 0.0256 \\
ed25519 \cite{eddsa}                              & 253 & sha256 & 0.0122  & 0.0116 & 0.0227 & 0.0126  & 0.0117 & 0.024  \\
jubjub                               & 252 & sha256 & 0.0131  & 0.0123 & 0.0253 & 0.0127  & 0.0142 & 0.0352 \\
e222 \cite{e521} & 220 & sha224 & 0.0123 & 0.0096 & 0.0206 & 0.0150 & 0.0089 & 0.0168 \\
\hline
\end{tabular}
\label{tab:edwards}
\end{table}

Table \ref{tab:koblitz} presents the performance comparison of Koblitz curves for EdDSA and ECDSA across various elliptic curve parameters. The table includes key performance metrics such as key generation, signing, and verification times for each curve. The curves are characterized by their form, name, bit-length n, and the hash function used. Generally, larger bit-lengths result in longer processing times due to increased cryptographic complexity. For both EdDSA and ECDSA, key generation times remain relatively consistent across curves, while signing and verification times vary more significantly. Higher bit-length curves like b571 and k571 exhibit the longest processing times, whereas smaller curves such as sect113r1 and wap-wsg-idm-ecid-wtls1 demonstrate the fastest performance. The comparison highlights that EdDSA typically has faster signing operations, whereas ECDSA shows more variation in signing and verification times. These insights are valuable for selecting the appropriate Koblitz curve based on the required balance between security and computational efficiency.

\begin{table}[H]
\centering
\caption{Performance of Binary Field Koblitz Form for ECDSA and EdDSA}
\begin{tabular}{lll|lll|lll}
\hline
 & & &  & \textbf{EdDSA} & & & \textbf{ECDSA} & \\
 \hline
\textbf{Curve} & \textbf{n} & \textbf{Hash} & \textbf{KeyGen (s)} & \textbf{Sign (s)} & \textbf{Verify (s)} & \textbf{KeyGen (s)} & \textbf{Sign (s)} & \textbf{Verify (s)} \\
\hline
b571                   & 570 & sha512 & 8.194  & 3.7683 & 17.4123 & 8.216  & 8.3472 & 16.174  \\
k571                   & 570 & sha512 & 8.2713 & 3.9519 & 16.1562 & 8.4748 & 8.2457 & 17.1495 \\
c2tnb431r1             & 418 & sha512 & 3.386  & 2.1913 & 7.4535  & 3.3694 & 3.374  & 6.7825  \\
b409                   & 409 & sha512 & 3.4035 & 1.9555 & 6.1243  & 2.9679 & 3.005  & 6.114   \\
k409                   & 407 & sha512 & 2.9362 & 2.0002 & 6.2678  & 3.0423 & 3.2321 & 6.0597  \\
c2pnb368w1             & 353 & sha384 & 1.9927 & 1.5442 & 4.3315  & 2.039  & 2.2011 & 4.2777  \\
c2tnb359v1             & 353 & sha384 & 2.0163 & 1.4565 & 3.9728  & 1.9861 & 2.0191 & 3.9461  \\
c2pnb304w1             & 289 & sha384 & 1.3413 & 1.2259 & 2.405   & 1.2239 & 1.1946 & 2.6447  \\
b283                   & 282 & sha384 & 1.0931 & 0.9905 & 2.3768  & 1.0412 & 1.0213 & 2.0634  \\
k283                   & 281 & sha384 & 1.2124 & 0.9356 & 2.0528  & 0.9805 & 0.9449 & 2.0996  \\
c2pnb272w1             & 257 & sha384 & 0.7865 & 0.8815 & 1.709   & 1.0488 & 0.8912 & 1.7936  \\
ansit239k1             & 238 & sha256 & 0.6724 & 0.6275 & 1.2682  & 0.7017 & 0.888  & 1.2746  \\
c2tnb239v1             & 238 & sha256 & 0.7212 & 0.7171 & 1.3736  & 0.6199 & 0.6499 & 1.189   \\
c2tnb239v2             & 237 & sha256 & 0.6987 & 0.6232 & 1.234   & 0.6207 & 0.5708 & 1.2449  \\
c2tnb239v3             & 236 & sha256 & 0.6312 & 0.6157 & 1.2617  & 0.6048 & 0.593  & 1.1998  \\
b233                   & 233 & sha256 & 0.5879 & 0.5479 & 1.2912  & 0.7237 & 0.8425 & 1.1725  \\
k233                   & 232 & sha256 & 0.6677 & 0.5965 & 1.6555  & 0.6128 & 0.5404 & 1.1134  \\
ansit193r1             & 193 & sha224 & 0.4117 & 0.3504 & 0.6917  & 0.322  & 0.3276 & 0.7053  \\
ansit193r2             & 193 & sha224 & 0.3241 & 0.3212 & 0.7989  & 0.3586 & 0.3558 & 0.6204  \\
c2pnb208w1             & 193 & sha224 & 0.3913 & 0.3725 & 0.7813  & 0.374  & 0.3936 & 0.7869  \\
c2tnb191v1             & 191 & sha224 & 0.3042 & 0.3416 & 0.7359  & 0.3341 & 0.3189 & 0.6884  \\
c2tnb191v2             & 190 & sha224 & 0.3182 & 0.327  & 0.6551  & 0.3281 & 0.3544 & 0.6656  \\
c2tnb191v3             & 189 & sha224 & 0.327  & 0.3333 & 0.7207  & 0.3449 & 0.322  & 0.6343  \\
b163                   & 163 & sha224 & 0.1969 & 0.2011 & 0.4037  & 0.2064 & 0.2351 & 0.5114  \\
c2pnb163v1             & 163 & sha224 & 0.2132 & 0.2122 & 0.4019  & 0.2335 & 0.2039 & 0.406   \\
k163                   & 163 & sha224 & 0.2729 & 0.2496 & 0.5028  & 0.2318 & 0.2229 & 0.4173  \\
ansit163r1             & 162 & sha224 & 0.2265 & 0.2332 & 0.4603  & 0.2074 & 0.2169 & 0.4184  \\
c2pnb163v2             & 162 & sha224 & 0.2322 & 0.2764 & 0.3987  & 0.2134 & 0.2638 & 0.4357  \\
c2pnb163v3             & 162 & sha224 & 0.228  & 0.2051 & 0.4114  & 0.2158 & 0.1913 & 0.3957  \\
c2pnb176w1             & 161 & sha224 & 0.2576 & 0.2598 & 0.4752  & 0.222  & 0.2353 & 0.5246  \\
sect131r1              & 131 & sha160   & 0.1067 & 0.1088 & 0.2342  & 0.1047 & 0.1085 & 0.2301  \\
sect131r2              & 131 & sha160   & 0.1174 & 0.0992 & 0.2317  & 0.1075 & 0.1073 & 0.2114  \\
sect113r1              & 113 & sha160   & 0.0821 & 0.0799 & 0.1372  & 0.0734 & 0.0688 & 0.145   \\
sect113r2              & 113 & sha160   & 0.0693 & 0.075  & 0.1523  & 0.069  & 0.0805 & 0.1484  \\
k113 & 112 & sha160   & 0.0795 & 0.0759 & 0.1487  & 0.1032 & 0.0828 & 0.1694 \\
\hline
\end{tabular}
\label{tab:koblitz}
\end{table}

The table \ref{tab:weierstass1} and \ref{tab:weierstass2} present the performance metrics of various Weierstrass curve implementations used for the EdDSA and ECDSA elliptic curve signature algorithms. The table includes details on the curve name, the order of the curve denoted by n, the hash function used, and the performance timings for key generation, signing, and verification operations. Each row corresponds to a specific curve configuration with performance measurements provided for both EdDSA and ECDSA, where time is recorded in seconds. For EdDSA, the key generation times range from approximately 0.006 to 0.077 seconds, signing times from 0.006 to 0.015 seconds, and verification times from 0.013 to 0.120 seconds. Similarly, ECDSA shows key generation times from 0.007 to 0.077 seconds, signing times from 0.0058 to 0.0699 seconds, and verification times from 0.0132 to 0.1278 seconds. This data allows for a detailed comparison of the efficiency of different elliptic curves for these cryptographic operations, aiding in the selection of the most optimal curve for specific use cases.

\begin{table}[H]
\centering
\caption{Performance of Prime Field Weierstrass Form for ECDSA and EdDSA \#1}
\begin{tabular}{lll|lll|lll}
\hline
 & & &  & \textbf{EdDSA} & & & \textbf{ECDSA} & \\
 \hline
\textbf{Curve} & \textbf{n} & \textbf{Hash} & \textbf{KeyGen (s)} & \textbf{Sign (s)} & \textbf{Verify (s)} & \textbf{KeyGen (s)} & \textbf{Sign (s)} & \textbf{Verify (s)} \\
\hline
bn638 & 638 & sha512 & 0.0692 & 0.0248 & 0.1205 & 0.0771 & 0.0636 & 0.1278 \\
bn606 & 606 & sha512 & 0.053 & 0.0222 & 0.1039 & 0.0518 & 0.0699 & 0.1062 \\
bn574 & 574 & sha512 & 0.0444 & 0.0283 & 0.106 & 0.0476 & 0.0452 & 0.1006 \\
bn542 & 542 & sha512 & 0.0392 & 0.0184 & 0.0809 & 0.0418 & 0.0409 & 0.0868 \\
p521 & 521 & sha512 & 0.0734 & 0.0188 & 0.0781 & 0.0372 & 0.0429 & 0.076 \\
brainpoolp512r1 & 512 & sha512 & 0.0436 & 0.0263 & 0.0897 & 0.0341 & 0.0347 & 0.0684 \\
brainpoolp512t1 & 512 & sha512 & 0.0462 & 0.0182 & 0.0698 & 0.04 & 0.0351 & 0.0761 \\
fp512bn \cite{fpbn} & 512 & sha512 & 0.036 & 0.0175 & 0.0715 & 0.0348 & 0.0336 & 0.0655 \\
numsp512d1 & 512 & sha512 & 0.0482 & 0.0234 & 0.0875 & 0.0413 & 0.0356 & 0.0678 \\
eccfrog512ck2 \cite{eccfrog512ck2} & 512 & sha512 & 0.0421 & 0.0205 & 0.0961 & 0.0503 & 0.0526 & 0.0734 \\
gost512 & 511 & sha512 & 0.0353 & 0.0169 & 0.0745 & 0.0349 & 0.0327 & 0.0721 \\
bn510 & 510 & sha512 & 0.03 & 0.0165 & 0.0627 & 0.0396 & 0.0354 & 0.0784 \\
bn478 & 478 & sha512 & 0.0274 & 0.0151 & 0.0545 & 0.0302 & 0.0301 & 0.0657 \\
bn446 & 446 & sha512 & 0.023 & 0.0164 & 0.0541 & 0.0295 & 0.0234 & 0.0463 \\
bls12-638 \cite{curve22103} \cite{BLS12-638} & 427 & sha512 & 0.0424 & 0.0241 & 0.0838 & 0.0495 & 0.0402 & 0.0754 \\
bn414 & 414 & sha512 & 0.0234 & 0.0136 & 0.0391 & 0.0208 & 0.0206 & 0.0439 \\
brainpoolp384r1 & 384 & sha384 & 0.0176 & 0.0118 & 0.0342 & 0.0176 & 0.0171 & 0.0355 \\
brainpoolp384t1 & 384 & sha384 & 0.0183 & 0.0124 & 0.0487 & 0.0183 & 0.0177 & 0.0373 \\
fp384bn \cite{fpbn} & 384 & sha384 & 0.0165 & 0.0109 & 0.0342 & 0.0189 & 0.0174 & 0.0345 \\
numsp384d1 & 384 & sha384 & 0.0276 & 0.0121 & 0.0389 & 0.0185 & 0.0175 & 0.0354 \\
p384 & 384 & sha384 & 0.0193 & 0.0128 & 0.0372 & 0.0174 & 0.0171 & 0.0352 \\
bls24-477 \cite{curve22103} \cite{BLS12-638} & 383 & sha384 & 0.0241 & 0.0156 & 0.0489 & 0.0241 & 0.0216 & 0.0542 \\
bn382 & 382 & sha384 & 0.0173 & 0.0111 & 0.0524 & 0.027 & 0.0169 & 0.0334 \\
curve67254 \cite{curve22103} & 380 & sha384 & 0.0167 & 0.0113 & 0.0358 & 0.0184 & 0.0177 & 0.0497 \\
bn350 & 350 & sha384 & 0.0154 & 0.0095 & 0.0277 & 0.0136 & 0.0127 & 0.03 \\
brainpoolp320r1 & 320 & sha384 & 0.0169 & 0.0145 & 0.0237 & 0.0118 & 0.0111 & 0.022 \\
brainpoolp320t1 & 320 & sha384 & 0.0107 & 0.0104 & 0.0226 & 0.0124 & 0.0115 & 0.0235 \\
bn318 & 318 & sha384 & 0.0112 & 0.0087 & 0.0239 & 0.0118 & 0.0103 & 0.0219 \\
bls12-455 & 305 & sha384 & 0.0174 & 0.0146 & 0.0347 & 0.0194 & 0.0185 & 0.0465 \\
bls12-446 \cite{curve22103} & 299 & sha384 & 0.0172 & 0.0136 & 0.0312 & 0.016 & 0.0153 & 0.0318 \\
bn286 & 286 & sha384 & 0.0082 & 0.0073 & 0.0166 & 0.0085 & 0.0086 & 0.0171 \\
brainpoolp256r1 & 256 & sha256 & 0.0162 & 0.0125 & 0.0135 & 0.0068 & 0.0066 & 0.0143 \\
brainpoolp256t1 & 256 & sha256 & 0.0072 & 0.007 & 0.0137 & 0.0069 & 0.0086 & 0.0163 \\
fp256bn \cite{fpbn} & 256 & sha256 & 0.0066 & 0.0065 & 0.013 & 0.0068 & 0.0062 & 0.0129 \\
gost256 & 256 & sha256 & 0.0066 & 0.0063 & 0.014 & 0.0068 & 0.0096 & 0.0197 \\
numsp256d1 & 256 & sha256 & 0.0077 & 0.007 & 0.0207 & 0.007 & 0.0067 & 0.0144 \\
p256 & 256 & sha256 & 0.0115 & 0.0094 & 0.0137 & 0.0073 & 0.0064 & 0.0132 \\
secp256k1 & 256 & sha256 & 0.0067 & 0.0077 & 0.0293 & 0.0067 & 0.0066 & 0.0136 \\
tom256 \cite{tom256} & 256 & sha256 & 0.0139 & 0.007 & 0.0141 & 0.0077 & 0.0068 & 0.013 \\
bls12-381 \cite{BLS12-381} & 255 & sha256 & 0.0117 & 0.0112 & 0.0227 & 0.0179 & 0.0112 & 0.0237 \\
pallas \cite{pastacurves} & 255 & sha256 & 0.0071 & 0.0063 & 0.0133 & 0.0063 & 0.0064 & 0.0133 \\
tweedledee \cite{tweedledum} & 255 & sha256 & 0.0138 & 0.0151 & 0.0193 & 0.0063 & 0.0062 & 0.0122 \\
tweedledum \cite{tweedledum} & 255 & sha256 & 0.007 & 0.0068 & 0.013 & 0.0068 & 0.006 & 0.0125 \\
vesta \cite{pastacurves} & 255 & sha256 & 0.0142 & 0.012 & 0.0132 & 0.0065 & 0.0066 & 0.0126 \\
\hline
\end{tabular}
\label{tab:weierstass1}
\end{table}

\begin{table}[H]
\centering
\caption{Performance of Prime Field Weierstrass Form for ECDSA and EdDSA \#2}
\begin{tabular}{lll|lll|lll}
\hline
 & & &  & \textbf{EdDSA} & & & \textbf{ECDSA} & \\
 \hline
\textbf{Curve} & \textbf{n} & \textbf{Hash} & \textbf{KeyGen (s)} & \textbf{Sign (s)} & \textbf{Verify (s)} & \textbf{KeyGen (s)} & \textbf{Sign (s)} & \textbf{Verify (s)} \\
\hline
bn254 & 254 & sha256 & 0.0064 & 0.006 & 0.0122 & 0.0162 & 0.0089 & 0.0144 \\
fp254bna \cite{Fp254BNa} & 254 & sha256 & 0.0068 & 0.0064 & 0.013 & 0.0065 & 0.0061 & 0.0131 \\
fp254bnb \cite{Fp254BNb} & 254 & sha256 & 0.007 & 0.0075 & 0.0133 & 0.008 & 0.0065 & 0.0138 \\
bls12-377 \cite{BLS12-377} & 253 & sha256 & 0.0123 & 0.012 & 0.0217 & 0.0128 & 0.0129 & 0.0238 \\
curve1174 \cite{curve1174} & 249 & sha256 & 0.0066 & 0.0062 & 0.0135 & 0.0068 & 0.0064 & 0.0134 \\
mnt4 & 240 & sha256 & 0.0064 & 0.0103 & 0.0264 & 0.006 & 0.0059 & 0.018 \\
mnt5-1 & 240 & sha256 & 0.0084 & 0.0105 & 0.0254 & 0.0086 & 0.0058 & 0.0173 \\
mnt5-2 & 240 & sha256 & 0.0146 & 0.0094 & 0.0132 & 0.0062 & 0.0058 & 0.0119 \\
mnt5-3 & 240 & sha256 & 0.0059 & 0.0055 & 0.0116 & 0.0059 & 0.0059 & 0.0114 \\
prime239v1 & 239 & sha256 & 0.0078 & 0.012 & 0.0203 & 0.0058 & 0.0055 & 0.0114 \\
prime239v2 & 239 & sha256 & 0.0064 & 0.006 & 0.0118 & 0.0059 & 0.0064 & 0.0119 \\
prime239v3 & 239 & sha256 & 0.006 & 0.0059 & 0.0121 & 0.0059 & 0.0057 & 0.0109 \\
secp224k1 & 225 & sha256 & 0.0053 & 0.0049 & 0.0105 & 0.0049 & 0.0048 & 0.0097 \\
brainpoolp224r1 & 224 & sha224 & 0.0051 & 0.0049 & 0.0097 & 0.005 & 0.0049 & 0.0115 \\
brainpoolp224t1 & 224 & sha224 & 0.0052 & 0.0049 & 0.0096 & 0.005 & 0.0062 & 0.0103 \\
curve4417 \cite{curve22103} & 224 & sha224 & 0.0063 & 0.0051 & 0.0107 & 0.0056 & 0.0048 & 0.0099 \\
fp224bn \cite{fpbn} & 224 & sha224 & 0.0051 & 0.0053 & 0.0103 & 0.0059 & 0.0049 & 0.0102 \\
p224 & 224 & sha224 & 0.0056 & 0.0109 & 0.014 & 0.0052 & 0.0052 & 0.0099 \\
bn222 & 222 & sha224 & 0.0049 & 0.0056 & 0.0096 & 0.005 & 0.0047 & 0.014 \\
curve22103 \cite{curve22103} & 218 & sha224 & 0.0052 & 0.005 & 0.0114 & 0.0053 & 0.0048 & 0.0099 \\
brainpoolp192r1 & 192 & sha224 & 0.0036 & 0.0033 & 0.0074 & 0.0047 & 0.0054 & 0.0119 \\
brainpoolp192t1 & 192 & sha224 & 0.0036 & 0.0033 & 0.0069 & 0.0041 & 0.0034 & 0.0074 \\
p192 & 192 & sha224 & 0.0039 & 0.0042 & 0.0099 & 0.0038 & 0.0037 & 0.0071 \\
prime192v2 & 192 & sha224 & 0.0038 & 0.0036 & 0.0071 & 0.0041 & 0.0036 & 0.0074 \\
prime192v3 & 192 & sha224 & 0.0038 & 0.0073 & 0.0074 & 0.004 & 0.0036 & 0.0074 \\
secp192k1 & 192 & sha224 & 0.0046 & 0.0036 & 0.011 & 0.0036 & 0.0041 & 0.007 \\
bn190 & 190 & sha224 & 0.0035 & 0.0032 & 0.0062 & 0.0035 & 0.0033 & 0.0071 \\
secp160k1 & 161 & sha224 & 0.0043 & 0.0059 & 0.0092 & 0.0056 & 0.0042 & 0.0058 \\
secp160r1 & 161 & sha224 & 0.004 & 0.0029 & 0.0053 & 0.0026 & 0.0022 & 0.0049 \\
secp160r2 & 161 & sha224 & 0.0038 & 0.0039 & 0.0063 & 0.0025 & 0.0024 & 0.0048 \\
brainpoolp160r1 & 160 & sha224 & 0.0039 & 0.0034 & 0.0049 & 0.0026 & 0.0024 & 0.0051 \\
brainpoolp160t1 & 160 & sha224 & 0.003 & 0.0025 & 0.0055 & 0.003 & 0.0024 & 0.0053 \\
mnt3-1 & 160 & sha224 & 0.0027 & 0.0024 & 0.0048 & 0.0026 & 0.0025 & 0.005 \\
mnt3-2 & 160 & sha224 & 0.0025 & 0.0028 & 0.0049 & 0.0025 & 0.0026 & 0.0049 \\
mnt3-3 & 160 & sha224 & 0.0025 & 0.0024 & 0.0052 & 0.0026 & 0.0025 & 0.0052 \\
mnt2-1 & 159 & sha160 & 0.0027 & 0.0023 & 0.0064 & 0.005 & 0.0043 & 0.0061 \\
mnt2-2 & 159 & sha160 & 0.0034 & 0.0025 & 0.0063 & 0.0028 & 0.0024 & 0.0052 \\
bn158 & 158 & sha160 & 0.0026 & 0.0022 & 0.0045 & 0.0034 & 0.0023 & 0.0048 \\
mnt1 & 156 & sha160 & 0.0029 & 0.0025 & 0.0053 & 0.0027 & 0.0025 & 0.008 \\
secp128r1 & 128 & sha160 & 0.0019 & 0.0016 & 0.0029 & 0.0017 & 0.0019 & 0.0042 \\
secp128r2 & 126 & sha160 & 0.0016 & 0.0014 & 0.0032 & 0.0028 & 0.0036 & 0.0075 \\
secp112r1 & 112 & sha160 & 0.0014 & 0.0012 & 0.0023 & 0.0018 & 0.0012 & 0.0024 \\
secp112r2 & 110 & sha160 & 0.0013 & 0.0011 & 0.0028 & 0.0013 & 0.0011 & 0.0029 \\
\hline
\end{tabular}
\label{tab:weierstass2}
\end{table}

The performance of Edwards, Koblitz, and Weierstrass curves for cryptographic operations is compared across key generation, signing, and verification times. Edwards curves, particularly in the form of EdDSA, offer efficient performance with near-linear scaling and fast execution, especially for smaller curves like ed25519. Koblitz curves, known for their use in binary fields for custom hardware design generally show competitive performance with moderate execution times. Weierstrass curves, typically used in traditional ECDSA, exhibit slower scaling compared to Edwards and Koblitz, with more noticeable increases in execution times as the curve size grows. Overall, Edwards curves provide the best performance for modern applications requiring both speed and security, followed by Koblitz and Weierstrass curves.

\subsection{High Level Comparison}

In the previous section, we analyzed digital signature algorithms (DSAs) independently across different configurations. Here, we compare these algorithms against each other.

The table \ref{tab:nist} summarizes NIST's recommendations for key sizes across different cryptographic algorithms, providing the equivalent key sizes for symmetric encryption, RSA, DSA, and ECC, along with the expected lifetime of the security.

\begin{minipage}{\linewidth} 
\begin{table}[H]
\centering
\caption{NIST's Key Size Recommendations}
\begin{tabular}{llll}
\toprule
\textbf{Security Level} & \textbf{RSA \& DSA Key Size} & \textbf{ECC Key Size} & \textbf{Expected Lifetime} \\
\midrule
80 & 1024 & 160 & Until 2010 \\
112 & 2048 & 224 & Until 2030 \\
128 & 3072 & 256 & Beyond 2030 \\
192 & 7680 & 384 & Much Beyond 2030 \\
256 & 15360 & 521 & Far Beyond 2030 \\
\bottomrule
\end{tabular}
\label{tab:nist}
\end{table}
\end{minipage}
\newline
\newline

By default, both ECDSA and EdDSA provide a 256-bit security level. Therefore, we compare 256-bit ECDSA and EdDSA with 3072-bit RSA and DSA, as all three configurations offer an equivalent 128-bit security level. For a 192-bit security level, ECDSA and EdDSA require 384-bit keys whereas RSA and DSA require 7680-bit keys. Thus, we use the P-384 curve in the Weierstrass form for ECDSA and the Numsp384t1 curve in the Edwards form for EdDSA. To achieve a 256-bit security level, ECDSA and EdDSA require 521-bit keys whereas RSA and DSA requires 15360-bit keys. Therefore, we adopt the P-521 curve in the Weierstrass form for ECDSA and the E-521 curve in the Edwards form for EdDSA.

The performance of RSA, DSA, ECDSA, and EdDSA is evaluated at different security levels—128-bit, 192-bit, and 256-bit—based on key generation, signing, and verification times. The results are summarized in Table \ref{tab:default_dsas}, Table \ref{tab:default_dsas_192}, and Table \ref{tab:default_dsas_256}.

\begin{table}[H]
\centering
\caption{Performance of Cryptosystems with Default Configurations for 128-bit Security Level}
\begin{tabular}{llllllll}
\hline
\textbf{Algorithm} & \textbf{Form} & \textbf{Curve} & \textbf{Key Size} & \textbf{Hash} & \textbf{KeyGen (s)} & \textbf{Sign (s)} & \textbf{Verify (s)} \\
\hline
RSA   & - & - & 3072 & sha256 & 9.3434 & 0.0704 & 0.0693 \\
DSA   & - & - & 3072 & sha256 & 41.2033 & 0.0065 & 0.0126 \\
ECDSA & Weierstrass & secp256k1 & 256  & sha256 & 0.0067 & 0.0066 & 0.0136 \\
ECDSA & Weierstrass & p256 & 256 & sha256 &   0.0073 & 0.0064 & 0.0132 \\
EdDSA & Edwards & ed25519 & 253  & sha256 & 0.0122 & 0.0116 & 0.0227 \\
\hline
\end{tabular}
\label{tab:default_dsas}
\end{table}

\begin{table}[h!]
\centering
\caption{Performance of Cryptosystems with Default Configurations for 192-bit Security Level}
\begin{tabular}{llllllll}
\hline
\textbf{Algorithm} & \textbf{Form} & \textbf{Curve} & \textbf{Key Size} & \textbf{Hash} & \textbf{KeyGen (s)} & \textbf{Sign (s)} & \textbf{Verify (s)} \\
\hline
RSA & - & - & 7680  & sha384 & 43.8724  & 0.9308 & 0.9995 \\
DSA & - & - & 7680  & sha384 & 432.1513 & 0.0412 & 0.0833 \\
ECDSA & Weierstrass & p384 & 384 & sha384 & 0.0174 & 0.0171 & 0.0352 \\
EdDSA & Edwards & numsp384t1 & 382 & sha384 & 0.0356 & 0.0201 & 0.0616 \\
\hline
\end{tabular}
\label{tab:default_dsas_192}
\end{table}

\begin{table}[h!]
\centering
\caption{Performance of Cryptosystems with Default Configurations for 256-bit Security Level}
\begin{tabular}{llllllll}
\hline
\textbf{Algorithm} & \textbf{Form} & \textbf{Curve} & \textbf{Key Size} & \textbf{Hash} & \textbf{KeyGen (s)} & \textbf{Sign (s)} & \textbf{Verify (s)} \\
\hline
RSA   & - & - & 15360 & sha512 & 455.00 & 7.0423 & 7.6164 \\
DSA   & - & - & 15360 & sha512 & 7172.1030& 0.1937 & 0.3868 \\
ECDSA & Weierstrass & p521 & 521 & sha512 & 0.0372 & 0.0429 & 0.076 \\
EdDSA & Edwards & e521 & 519 & sha512 & 0.0667 & 0.032 & 0.1357 \\
\hline
\end{tabular}
\label{tab:default_dsas_256}
\end{table}

At the 128-bit security level, ECDSA (secp256k1 \& p256 or secp256r1) and EdDSA (ed25519) outperform RSA and DSA (3072-bit) in all metrics. RSA key generation takes over 9 seconds, while DSA takes just over 41 seconds. In contrast, both ECDSA and EdDSA generate keys in milliseconds. Signing and verification are also significantly faster with ECDSA and EdDSA. Among elliptic curve algorithms, ECDSA has a slight edge in signing and verification speed, making it the most efficient option at this level.

For 192-bit security, ECDSA (P-384) and EdDSA (Numsp384t1) continue to be superior to RSA (7680-bit) and DSA. RSA's key generation time reaches over 43 seconds, while DSA's is even higher at 432 seconds, making both impractical for dynamic key creation. In contrast, ECDSA and EdDSA generate keys in under 40 milliseconds, with much faster signing and verification times as well.

At the 256-bit security level, RSA and DSA (15360-bit) become even less practical due to extremely long key generation times of 455 seconds and over 7000 seconds, respectively, and slow signing and verification speeds. ECDSA (P-521) and EdDSA (E-521) are significantly faster, with ECDSA providing better verification performance, while EdDSA is slightly more efficient in signing.

So, we can summarize our observations as:

\begin{itemize}
    \item RSA consistently underperforms, particularly in key generation, making it impractical for dynamic use.
    \item DSA has faster signing and verification than RSA at all levels, but suffers from extremely slow key generation, especially as security level increases.
    \item ECDSA is generally faster in verification, while EdDSA is more efficient for signing across different security levels.
    \item Both ECDSA and EdDSA significantly outperform RSA and DSA, making them the preferred options for modern cryptographic applications.
    \item As the security level increases, RSA's computation time grows exponentially, making it increasingly inefficient. DSA also exhibits very slow key generation times. In contrast, both ECDSA and EdDSA exhibit only linear growth, maintaining practical performance even at higher security levels.
\end{itemize}

\section{Conclusion}

In this paper, we have compared the performance of three prominent digital signature algorithms (DSAs)—RSA, DSA, ECDSA, and EdDSA—across various security levels (128-bit, 192-bit, and 256-bit). The analysis was conducted by evaluating key generation, signing, and verification times for each algorithm, using NIST's recommended key sizes for equivalent security levels.

Our results clearly demonstrate that both ECDSA and EdDSA outperform RSA and DSA in terms of speed and efficiency, particularly in the context of dynamic key generation. RSA’s and DSA's key generation times increase significantly with larger key sizes, making it less suitable for environments where quick key creation and signing are critical. In contrast, ECDSA and EdDSA, both based on elliptic curve cryptography, maintain fast performance across all security levels, with EdDSA being more efficient in signing operations and ECDSA excelling in verification.

At the 128-bit security level, ECDSA (secp256k1) and EdDSA (ed25519) exhibit superior performance over RSA and DSA (3072-bit) in all metrics, including key generation, signing, and verification. As we increase the security level to 192-bit and 256-bit, RSA’s performance continues to degrade exponentially, DSA time outs while the elliptic curve algorithms scale much more efficiently.

All experiments were conducted on a machine equipped with an 11th Gen Intel Core TM i7-11370H processor, running at 3.30 GHz with 8 cores. This CPU, part of Intel's Tiger Lake family, is commonly found in high-performance laptops. Providing these details ensures transparency about the computational environment used for the experiments.

This study highlights the increasing practicality of elliptic curve-based algorithms like ECDSA and EdDSA as security requirements grow. For modern cryptographic applications, particularly those demanding high-performance and scalability, ECDSA and EdDSA provide a clear advantage over RSA and DSA. Therefore, we recommend adopting ECDSA and EdDSA for applications requiring secure, fast, and efficient digital signatures.

\newpage
\bibliographystyle{unsrt}

\end{document}